\documentclass{emulateapj}
\usepackage{apjfonts}

\usepackage{amssymb,natbib}
\def\beq{\begin{equation}}
\def\eeq{\end{equation}}
\begin{document}
\slugcomment{Astrophysical Journal, Accepted}
\shorttitle{DIFFERENTIALLY ROTATING STARS}
\shortauthors{MACGREGOR ET AL.}
\title{ON THE STRUCTURE AND PROPERTIES OF DIFFERENTIALLY ROTATING, MAIN-SEQUENCE STARS IN THE $1 - 2\ M_\sun$ RANGE} 
\author{K.~B.~MacGregor,
Stephen Jackson,
Andrew Skumanich,
T.~S.~Metcalfe}
\affil{\emph{High Altitude Observatory, NCAR,
P.~O.~Box 3000, Boulder, CO 80307\footnote
{The National Center for Atmospheric Research (NCAR)
is sponsored by the National Science Foundation.}}}
\begin{abstract}

We conduct a systematic examination of the properties of models for 
chemically homogeneous, differentially rotating, main-sequence stars of 
mass $1- 2\ M_\sun$.  The models were constructed using a code based on a 
reformulation of the self-consistent field method of computing the 
equilibrium stellar structure for a specified conservative internal 
rotation law.  The code has recently been upgraded with the addition of 
new opacity, equation of state, and energy generation routines, and a 
mixing-length treatment of convection in the outer layers of the stellar 
interior.  Relative to nonrotating stars of the same mass, these models 
all have reduced luminosities and effective temperatures, and flattened 
photospheric shapes (i.e., decreased polar radii) with equatorial radii 
that can be larger or smaller, depending on the degree of differential 
rotation.  For a fixed ratio of the axial rotation rate to the surface 
equatorial rotation rate, increasingly rapid rotation generally deepens 
convective envelopes, shrinks convective cores, and can lead to the 
presence of a convective core (envelope) in a $1\ M_\sun$ ($2\ M_\sun$) 
model, a feature that is absent in a nonrotating star of the same mass.  
The positions of differentially rotating models for a given mass $M$ in 
the H-R diagram can be shifted in such a way as to approximate the 
nonrotating ZAMS over ranges in luminosity and effective temperature that 
correspond to a mass interval between $M$ and about 0.7 $M$. We briefly 
note a few of the implications of these results, including (i) possible 
ambiguities arising from similarities between the properties of rotating 
and nonrotating models of different masses, (ii) a reduced radiative 
luminosity for a young, rapidly rotating Sun, (iii) the nuclear 
destruction of lithium and other light metallic species in the layers 
beneath an outer convective envelope, and (iv), the excitation of 
solar-like oscillations and the operation of a solar-like hydromagnetic 
dynamo in some $1.5 - 2\ M_\sun$ stars.

\end{abstract}
\section{Introduction}

Rotation is a universal stellar physical attribute.  This conclusion is 
supported by an enormous body of observational data, accrued over a period 
of nearly 100 years, from which information about rotational speeds and 
periods, the dependence of these quantities on parameters such as mass and 
age (i.e., evolutionary state), and the effects of rotation on the shape, 
effective temperature, chemical homogeneity, and other basic properties of 
stars have been inferred.  Despite increasing evidence that rotation can 
have a significant impact on a variety of stellar characteristics, it is 
generally not included as a component of the structural/evolutionary 
models which are the primary tools for the interpretation of observations. 
This omission is, in part, a consequence of the increased complexity of 
the problem of determining the structure and evolution of a rapidly, 
differentially rotating star: a one-dimensional model necessarily becomes 
two-dimensional, the gravitational potential must be derived by solution 
of Poisson's equation, and the uncertain physics of convective and 
circulatory flows, along with other rotation-dependent mechanisms that 
contribute to angular momentum redistribution and chemical mixing, needs 
to be addressed.

Some progress toward the development of a straightforward yet robust 
technique for computing the internal structure of a rotating star has 
recently been made with the implementation of a new version of the 
self-consistent field (SCF) method (Jackson, MacGregor, \& Skumanich 2005, 
hereafter Paper I).  In its original form (Ostriker \& Mark 1968; see also 
Jackson 1970), the SCF approach to computing a model for an assumed 
conservative law of rotation consisted of two separate steps: (i) 
determination of the gravitational potential for a given distribution of 
the mass density $\rho$, followed by (ii) solution of the equations of 
stellar structure with the potential from (i) to update the equilibrium 
distributions of $\rho$ and other quantities.  The process was initiated 
with a trial distribution for $\rho$, and the two steps were executed 
sequentially and iterated upon until the densities from (i) and (ii) 
agreed to within a specified tolerance.  While the SCF method was 
successfully used to construct detailed models of rotating 
upper-main-sequence stars (Bodenheimer 1971), it failed to converge for 
objects less massive than about 9 $M_\sun$, behavior which precluded its 
application to intermediate- and low-mass stars (see, e.g., Chambers 1976; 
Clement 1978, 1979; Paper I).

As described in Paper I (see also \S 2 below), the reformulated SCF method 
circumvents the difficulties responsible for nonconvergence in lower-mass, 
more centrally condensed stars with an approach that entails the 
specification of trial functions for the pressure $P$, the temperature 
$T$, and the shape of constant-density surfaces, together with iterative 
adjustment of both the profiles and central values of $P$ and $T$.  With 
these modifications, the method is capable of producing self-consistent 
models for rotating main-sequence stars of all masses.  It has been 
validated through detailed comparisons with stellar models (both rotating 
and nonrotating) for masses $\ge 2\ M_\sun$ computed by other 
investigators using alternative techniques (Paper I), and has been applied 
in an examination of the photospheric shape of the Be star Achernar, an 
object revealed by interferometric observations as highly flattened by 
rapid rotation (Jackson, MacGregor, \& Skumanich 2004, and references 
therein).  More recently, the structure code based on the method has 
undergone considerable renovation, with the replacement of routines for 
the equation of state, nuclear energy generation rates, and opacities, and 
the addition of a mixing length treatment of convection. It is now 
equipped for use in an investigation of the effects of rotation on the 
structure and properties of stars with masses $\le 2\ M_\sun$.

In the present paper, we use the formalism described above to conduct a 
detailed survey of the structural characteristics of differentially 
rotating stars having masses $M$ in the range $1 \le M \le 2\ M_\sun$. In 
the absence of rotation, this mass interval encompasses considerable 
variation in internal properties, with the gross structure of objects at 
the lower limit consisting of an inner radiative core and an outer 
convective envelope, changing to a convective core and radiative envelope 
for objects at the upper limit.  The influence of rotation on the basic 
morphology of stellar interiors for these masses has not received much 
attention, with rotation-related effects usually treated as perturbations 
to the nonrotating structure, if at all (see, e.g., Thompson et al. 2003). 
However, observations indicate the occurrence of surface rotation speeds 
rapid enough to imply non-negligible modifications to many stellar 
properties, particularly if the rotation is differential: the projected 
rotation speeds of 2 $M_\sun$ main-sequence stars are typically in excess 
of 100 km s$^{-1}$, and comparable values have been measured for near 
zero-age main-sequence (ZAMS) 1 $M_\sun$ stars in young clusters (Stauffer 
1991; Wolff \& Simon 1997; Tassoul 2000, and references therein).

Motivated by these considerations, we have constructed an extensive set of 
self-consistent models for $1-2\ M_\sun$ ZAMS stars, to systematically 
study the dependence of stellar characteristics on the rate and degree of 
differential rotation. Our SCF models extend previous computational 
results for stars in this mass range, some of which were obtained for 
uniformly rotating configurations (Faulkner, Roxburgh, \& Stritmatter 
1968; Sackmann 1970; Kippenhahn \& Thomas 1970; Papaloizou \& Whelan 1973; 
Roxburgh 2004), while others were obtained using either approximate 
methods for nonconservative differential rotation (Endal \& Sofia 1981; 
Pinsonneault et al. 1989; Eggenberger, Maeder, \& Meynet 2005) or a 
non-SCF, finite-difference technique in the case of conservative rotation 
(Clement 1979). In \S2 we provide a synopsis of the new SCF method, 
briefly describing its implementation in a code for computing the 
structure of chemically homogeneous, differentially rotating main sequence 
stars, the improvements and extensions to the input physics that have been 
made since Paper I, and the results of tests of code reliability through 
comparisons with extant models for nonrotating ZAMS stars with masses 
$\leq 2\ M_\sun$.  An examination of the properties of models spanning a 
wide range of internal rotation characteristics is presented in \S3, with 
particular attention paid to the behavior of such quantitites as 
luminosity and effective temperature, central thermodynamic properties, 
the location and extent of convective regions, and the size and shape of 
the stellar photosphere. Among the consequences of these rotation-induced 
changes in stellar properties is a shift in the position of objects in the 
classical H-R diagram (HRD); indeed, for some adopted rotation laws, the 
resulting structural modifications can enable a differentially rotating 
star to have the same $T_{eff}$ and $L$ values as a non-rotating star of 
significantly lower mass and to thereby occupy the same position in the 
HRD.  In the concluding section of the paper (\S4), we use the model 
results to consider how the changes brought on by rapid differential 
rotation might affect other features of main-sequence stars in this mass 
range, such as the radiative luminosity of the young Sun and associated 
effects on the planets, the abundance of lithium in the surface layers of 
mid-F dwarf stars, and the presence of strong, large-scale magnetic fields 
in some A stars.

\section{The New SCF Method}

\subsection{Implementation}

The new version of the SCF method (described in detail in Paper I) is an 
iterative scheme that is initialized by specifying (i) a pair of 
one-dimensional trial functions, one for the temperature distribution and 
one for the pressure distribution (each normalized by its central value 
and defined over a spatial range that is normalized by the equatorial 
radius of the star), and (ii) a two-dimensional normalized function 
describing the shape of the equidensity surfaces.  The normalized 
two-dimensional trial density distribution, which is used as the source 
term in Poisson's equation for the gravitational potential, follows from 
the equation of state and the trial functions for $P$ and $T$.  For a 
conservative law of rotation, in which the angular velocity depends only 
on the perpendicular distance from the axis of rotation, it is possible to 
define an effective potential from which both the gravitational and 
centrifugal forces can be derived.  The surfaces of constant effective 
potential (i.e., level surfaces) can then be identified and used to solve 
a set of ordinary differential equations analogous to the usual equations 
of stellar structure for nonrotating stars.  This solution step yields 
updated temperature and pressure distributions, allowing the iterative 
cycle to be repeated and the process continued until convergence is 
achieved. When the input and output functions representing the normalized 
temperature and pressure profiles are in agreement, the two parameters 
corresponding to the central temperature ($T_c$) and central pressure 
($P_c$) are adjusted by a Newton-Raphson technique to bring them closer to 
the actual physical conditions at the center of the final equilibrium 
model.  The entire procedure (consisting of an SCF loop nested inside a 
Newton-Raphson loop) is repeated until an acceptable level of agreement 
between the input and output values of the two central parameters is 
attained. When used to construct models for differentially rotating 
main-sequence stars, this reformulation of the SCF method has been shown 
to converge for all masses in the range $0.6 \leq M \leq 30\ M_\sun$, and 
for values of the dimensionless rotational kinetic energy $t$ (the ratio 
of the rotational kinetic energy to the absolute value of the 
gravitational potential energy of the configuration) as high as 0.10--0.12 
for intermediate- and high-mass models, and up to nearly 0.26 for some 
fully convective, highly flattened, disk-like, $1\ M_\sun$ models.  By 
comparison, the largest $t$ values among the non-SCF models computed by 
Clement (1978, 1979) were $\approx 0.18$ for 30 $M_\sun$ and $\le 0.12$ 
for models in the range $1.5 \le M \le 5 M_\sun$, whereas Bodenheimer 
(1971) obtained a 60 $M_\sun$ SCF model with $t \approx 0.24$.

As in Paper I (see also Jackson, MacGregor, \& Skumanich 2004), the 
internal rotation of each of the stellar models discussed in \S3 is given 
by an angular velocity distribution of the form
\beq
\Omega ( \varpi ) = {\Omega_0 \over 1 + \left( \alpha \varpi / R_e \right)^2 },
\eeq
where $\varpi = r\ {\rm sin}\ \theta$, and $R_e$ is the equatorial radius 
of the star.  For a given model, the constants $\Omega_0$ and $\alpha$ are 
prescribed parameters that characterize, respectively, the axial rotation 
rate and the ratio of the surface equatorial rotation rate to the axial 
rate (a measure of the degree of differential rotation), $\Omega_e / 
\Omega_0 = 1/(1 + \alpha^2)$.  In practice, the value of the first of 
these quantities is specified through the parameter $\eta = \Omega_0 / 
\Omega_{cr}$, with $\Omega_{cr}$ the equatorial angular velocity for which 
the magnitudes of the gravitational and centrifugal forces at $R_e$ are 
equal.

Rigorous answers to questions about the existence and uniqueness of 
solutions to complicated systems of integro-differential equations are 
generally very difficult to obtain.  Although specifying the two 
rotational parameters $\alpha$ and $\eta$ for chemically homogeneous 
models of fixed mass and composition does not always lead to a converged 
model, when it does, our experience with the code suggests that the model 
is unique.  For a given set of $(\alpha, \eta)$ values, using the SCF code 
to converge models of the same mass and chemical composition from two or 
more \emph{different} trial models always seems to lead to the \emph{same} 
final converged model.  On the other hand, it can be demonstrated that 
specifying the parameters $\alpha$ and $\Omega_0$ does not, in general, 
lead to a unique solution.  A minor drawback to use of the parameters 
($\alpha$, $\eta$) in presenting the results is that several important 
global properties, including the luminosity $L$, the total angular 
momentum $J$, and the dimensionless rotational kinetic energy $t$, are not 
monotonic functions of $\eta$ when $\alpha$ is held fixed (see also \S 
3.1).  Despite this shortcoming, we feel that the apparent uniqueness of 
the models corresponding to particular $(\alpha, \eta)$ values justifies 
the use of these quantities in presenting and discussing results.  We 
emphasize that sequences of models obtained by holding $\alpha$ fixed and 
varying $\eta$ should not be interpreted as any sort of evolutionary 
sequence.  The significance of models characterized by the same value of 
$\alpha$ is that they have the same degree of differential rotation, that 
is, the same $\Omega_e / \Omega_0$.  The ``half-width'' of the rotation 
profile, $\varpi_{1/2} = R_e / \alpha$, does, however, change from model 
to model for constant $\alpha$.

\subsection{Input Physics}

Since the publication of Paper I, the input physics for the SCF code 
described therein has been updated considerably, with the installation of 
software components that at various times were parts of the 
stellar-evolution code developed by Don VandenBerg at the University of 
Victoria.  All of the models presented in this paper were computed for the 
following abundances by weight of H, He, and heavy elements: $X = 0.7112$, 
$Y = 0.27$, and $Z = 0.0188$.  The opacities were obtained, as in 
VandenBerg et al.\ (2000), from tables of OPAL opacities calculated by 
Rogers \& Iglesias (1992) and from tables of low-temperature opacities 
calculated by Alexander \& Ferguson (1994), using interpolation 
subroutines written by VandenBerg (1983).  Other subroutines written by 
VandenBerg (1992) were utilized for the following: (i) the equation of 
state formulated by Eggleton, Faulkner, \& Flannery (1973, EFF); (ii) 
nuclear energy generation rates for hydrogen burning from Caughlan \& 
Fowler (1988), including the effect of electron screening as treated by 
Graboske et al.\ (1973) for the case of equilibrium abundances of CNO 
isotopes; and (iii) a standard mixing-length treatment of surface 
convective zones (see, e.g., Baker \& Temesvary 1966; Kippenhahn, Weigert, 
\& Hofmeister 1967).  We have made no attempt to incorporate any of the 
direct effects of rotation into the adopted convection model; instead, we 
have simply modified the nonrotating mixing-length description of 
convection by replacing the local gravitational acceleration, ${\bf g}$, 
with the effective gravitational acceleration, ${\bf g}_{eff} = {\bf g} - 
\Omega (\varpi)^2 \varpi\ {\bf e}_\varpi$ (i.e., ${\bf g}$ as reduced by 
the local centrifugal acceleration), averaged over equipotential surfaces.  
For all of the models, a value of 1.9 for the ratio of the mixing length 
to the pressure scale height has been adopted.

\subsection{Validation}

In view of the revisions and updates that have been made to the SCF code 
of Paper I, it seems worthwhile to make a careful comparison of our 
nonrotating models for stars on the lower main sequence with models for 
the same mass obtained from a standard (nonrotating) stellar evolution 
code For this purpose, we have used the current version of the 
evolutionary code of Christensen-Dalsgaard (1982, hereafter referred to as 
the JCD code) to generate two evenly spaced sequences of seven models 
each, spanning the mass range $0.8 \leq M \leq 2.0\ M_\sun$ along the 
ZAMS. These models have chemical composition $X = 0.711$, $Z = 0.019$, 
quite close to the abundances adopted for our SCF models, and were 
computed using the same value (1.9) of the mixing length parameter.  The 
basic EFF equation of state was used to construct one of the sequences, 
while the other was derived with the so-called CEFF equation of state, a 
modification of the EFF treatment that includes Coulomb corrections (see, 
e.g., Christensen-Dalsgaard \& Dappen 1992).  Aside from the inclusion or 
omission of Coulomb effects in the equation of state, the package of 
input-physics routines used to generate the JCD models is very similar to 
that installed in the SCF code (see, e.g., Christensen-Dalsgaard, 
Proffitt, \& Thompson 1993; Di Mauro \& Christensen-Dalsgaard 2001).  In 
addition to the intrinsic distinctions between the mathematical techniques 
used to obtain the two sets of models, the current version of the SCF code 
differs from the JCD code in the following ways: (i) the equation of state 
does not include treatment of Coulomb corrections; (ii) the photospheric 
pressure in the SCF code follows from the application of a different, 
simplified boundary condition (see Paper I); and (iii), there is a slight 
difference in the vintage of the nuclear energy generation data.

We have conducted a quantitative comparison of some of the important 
properties ($R$, $L$, $T_{eff}$, $P_c$, $T _c$, $\rho_c$) of SCF, JCD/EFF, 
and JCD/CEFF models for nonrotating ZAMS stars of the same mass.  A 
theoretical HRD indicating the positions of the various models is 
displayed in Figure 1.  The EFF and CEFF main sequences (the dashed and 
dotted lines, respectively) are very nearly coincident, with the former 
models shifted shifted along the locus relative to the latter models, 
toward somewhat lower luminosities and effective temperatures.  On the 
basis of the preceding discussion, we expect the SCF models to agree 
better with EFF models than with CEFF models, and inspection of the 
results plotted in Figure 1 indicate that this is generally the case.  
For the luminosity, the most sensitive of the stellar properties, the 
largest discrepancies occur for the $0.8\ M_\sun$ models, with the SCF $L$ 
about $2 \%$ lower than that of the EFF model of the same mass; this 
latter value is, in turn, about $7\%$ lower than $L$ for the corresponding 
CEFF model.  The luminosities of the SCF and EFF models are essentially 
equal for $2\ M_\sun$, while a comparison of the EFF and CEFF models for 
that mass reveals that the CEFF model is about $3\%$ more luminous. The 
effective temperatures and radii of the SCF models deviate from those of 
the corresponding EFF models by less than $1\%$, except for those models 
between 1.2 and 1.4 $M_\sun$ where the the magnitudes of the discrepancies 
are somewhat larger, $\approx 1\% - 2\%$.  The SCF values for $P_c$ and 
$\rho_c$ differ from the EFF values of those quantities by about $2\%$, 
with the relative difference in the $T_c$ values $\la 1\%$.  There is also 
good qualitative agreement between the SCF and JCD models with respect to 
the appearance or absence of convective cores and envelopes, and there is 
good quantitative agreement with respect to the radial extent and the 
enclosed mass of the principal radiative-convective interfaces.

\begin{figure}
\plotone{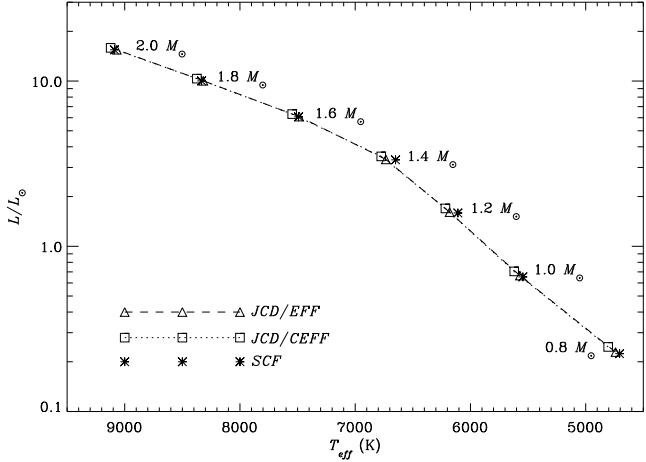}
\caption{A theoretical HR diagram showing the positions of nonrotating, 
chemically homogeneous stellar models for the indicated masses, as 
computed using different codes and equations of state.  The various 
symbols denote models obtained using the SCF code of the present paper 
($\ast$), a current version of the code of Christensen-Dalsgaard (1982, 
JCD) with the simplified equation of state of Eggleton, Faulkner, \& 
Flannery (1973, EFF) ($\vartriangle$), and the JCD code with a version of 
the EFF equation of state that includes Coulomb corrections (CEFF) 
($\sq$).  The dashed and dotted lines are the computed ZAMS locations 
derived from the JCD/EFF and JCD/CEFF models, respectively.}
\end{figure}

\section{Properties of Differentially Rotating Stellar Models}

\subsection{Convergence Properties}

The properties of SCF models for rotating ZAMS stars with masses $1 \leq M 
\leq 2\ M_\sun$ are summarized in Figures 2-8 and Table 1.  To facilitate 
discussion of the computed stellar characteristics, we adopt the 
convention, established in Paper I, of identifying each model by its mass, 
$M$, and the two rotational parameters, $\alpha$ and $\eta$, for the 
reasons discussed in \S 2.  Regions of the ($\alpha,\ \eta$) parameter 
space in which converged $6\ M_\sun$ SCF models can and cannot be obtained 
have been illustrated in Figure 3 of Paper I. While the ($\alpha$, $\eta$) 
planes for the lower-mass SCF models considered here closely resemble that 
for the $6\ M_\sun$ models, there are some important differences.  Of 
particular relevance to the present paper are the regions corresponding to 
Region II in Paper I, regions of relatively high angular momentum wherein 
the SCF method is incapable of producing converged models.  These 
\emph{forbidden zones} are considerably expanded for the 1 and 2 $M_\sun$ 
models: for the former models, the values ($\alpha_t,\ \eta_t$) 
corresponding to the lower tip of this region are $\alpha_t = 1.39$, 
$\eta_t \approx 2.4$, while for the latter, $\alpha_t = 2.83$ and $\eta_t 
\approx 6.5$.  Along each constant-$\alpha$ sequence for $0 < \alpha \le 
\alpha_t$, quantities such as the axial angular velocity $\Omega_0$ and 
the total angular momentum $J$ are non-monotonic functions of $\eta$, 
increasing to a maximum for $\eta \approx \eta_t$ and decreasing 
thereafter.  As $\alpha \rightarrow \alpha_t$, convergence in the vicinity 
of this maximum becomes significantly slower, and fails completely when 
$\alpha$ is large enough (i.e., $> \alpha_t$) to place the model within 
the forbidden zone.  For $\alpha_t < \alpha \leq 7$ (the largest value of 
$\alpha$ we have considered), converged models can be readily obtained on 
the low-$\eta$ sides of these regions. On the high-$\eta$ sides, converged 
models can be obtained only within the ranges $1.39 \le \alpha \le 3$ for 
the 1 $M_\sun$ models, and $2.83 \le \alpha \le 3.47$ for the 2 $M_\sun$ 
models.  The physical, mathematical, and computational reasons for the 
lack of convergence of models on either side of the forbidden zone are 
discussed in considerable detail in Paper I.  In this paper, we present 
results for complete constant-$\alpha$ sequences of models having $\alpha 
< \alpha_t$, but confine our attention to just the low-$\eta$ sides of 
forbidden zones for models with $\alpha > \alpha_t$.  Models on the 
high-$\eta$ sides of these regions have highly rotationally flattened, 
disk-like structures that will be the focus of a subsequent paper.

\begin{figure}
\plotone{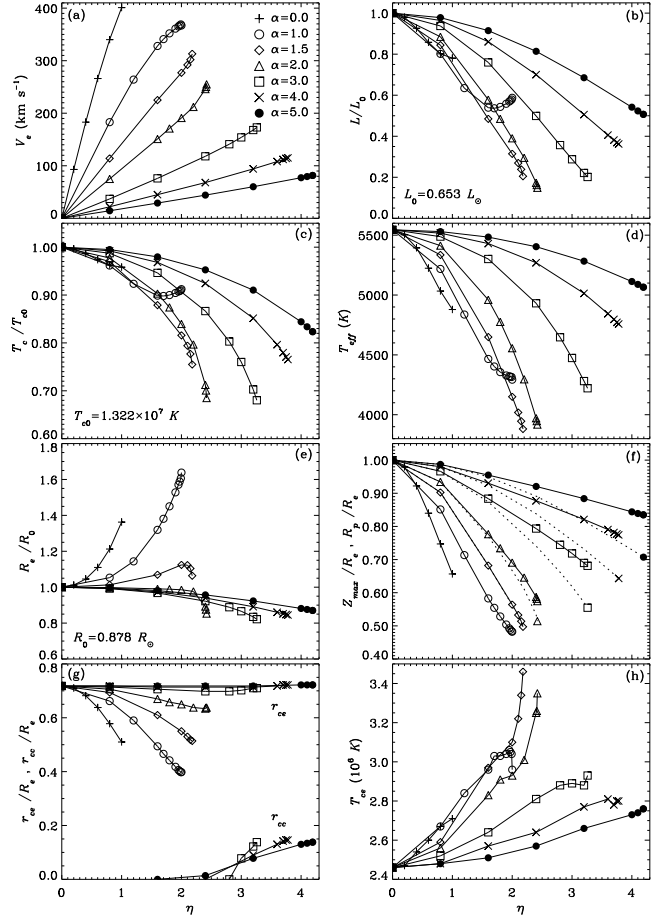}
\caption{Selected properties of differentially rotating, $1\ M_\sun$, ZAMS 
stellar models.  The model characteristics are shown as functions of 
$\eta$ for $0 \leq \alpha \leq 5$, where the parameters $\alpha$ and 
$\eta$ specify the assumed internal angular velocity distribution given by 
equation (1).  The quantities depicted in the various panels include: (a) 
the surface equatorial rotation speed $V_e$; (b) the luminosity $L$ in 
units of $L_0$, the luminosity of a nonrotating $1\ M_\sun$ model; (c) the 
central temperature $T_c$ relative to the corresponding value $T_{c0}$ for 
the nonrotating model; (d) the average effective temperature $T_{eff}$; 
(e) the equatorial radius $R_e$ in units of $R_0$, the radius of the 
nonrotating model; (f) the maximum perpendicular distance $Z_{max}$ from 
the equatorial plane to the photosphere (solid curves), and the polar 
radius $R_p$ (dotted curves), as fractions of $R_e$; (g) the radii of the 
base of the convective envelope $r_{ce}$ and the convective core $r_{cc}$, 
both measured in the equatorial plane relative to $R_e$; and, (h) the 
temperature $T_{ce}$ at the base of the convective envelope.}
\end{figure}

\begin{deluxetable*}{ccccc ccccc ccccc ccccc cc}
\tabletypesize{\small}
\tablecaption{Selected Models \label{t1}}
\label{tab:mod}
\tablewidth{\textwidth}
\tablehead{\colhead{Model} &
\colhead{I} &
\colhead{II} &
\colhead{III} &
\colhead{IV} &
\colhead{V} &
\colhead{VI} &
\colhead{VII} &
\colhead{VIII} &
\colhead{IX} &
\colhead{X} &
\colhead{XI} }
\startdata
$M$ &
         0.8 & 
           1 &
           1 &
           1 &
           1 &
         1.2 &
         1.2 &
         1.6 &
           2 &
           2 &
           2 \\
$\alpha$ &
           0 & 
 	   0 &
 	 1.5 &
	3.75 &
	   5 &
           0 &
           4 &
           0 &
	   0 &
	   3 &
	4.75 \\
$\eta$ &
           0 &
 	   0 &
 	1.55 &
	3.58 &
	4.15 &
           0 &
        3.74 &
           0 &
	   0 &
	5.64 &
         5.9 \\
$t$ &
           0 &
           0 &
       0.044 &
       0.079 &
       0.052 &
           0 &
       0.065 &
           0 &
           0 &
       0.047 &
       0.078 \\
$J$ &
           0 &
           0 &
        6.44 &
        7.27 &
        5.49 &
           0 &
        8.93 &
           0 &
           0 &
       23.71 &
       23.73 \\
$V_e$ &
           0 &
           0 &
         219 &
         123 &
          81 &
           0 &
         112 &
           0 &
           0 &
         205 &
         127 \\
$L$ &
       0.224 &
       0.653 &
       0.332 &
       0.224 &
       0.336 &
       1.591 &
       0.653 &
       6.075 &
      15.47  &
       7.560 &
       6.083 \\
$T_{\mbox{\emph{eff}}}$ &
     4710 &
     5540 &
     4710 &
     4700 &
     5080 &
     6110 &
     5550 &
     7490 &
     9090 &
     5890 &
     7480 \\
$\log\, g_s$ &
    4.635 &
    4.551 &
    4.478 &
    4.709 &
    4.679 &
    4.412 &
    4.616 &
    4.310 &
    4.337 &
    3.855 &
    4.404 \\
$R_e$ &
    0.714 &
    0.878 &
    0.934 &
    0.744 &
    0.767 &
    1.129 &
    0.904 &
    1.466 &
    1.590 &
    2.927 &
    1.512 \\
$R_p/R_e$ &
        1 &
        1 &
    0.697 &
    0.638 &
    0.712 &
        1 &
    0.668 &
        1 &
        1 &
    0.371 &
    0.549 \\
$Z_{\mbox{\emph{max}}}/R_e$ &
        1 &
        1 &
    0.697 &
    0.764 &
    0.837 &
        1 &
    0.791 &
        1 &
        1 &
    0.530 &
    0.770 \\
$\log\, P_c$ &
   17.078 &
   17.178 &
   17.100 &
   16.967 &
   17.019 &
   17.259 &
   17.101 &
   17.326 &
   17.287 &
   17.317 &
   17.282 \\
$\log\, T_c$ &
    7.046 &
    7.121 &
    7.069 &
    7.004 &
    7.039 &
    7.187 &
    7.094 &
    7.280 &
    7.326 &
    7.289 &
    7.255 \\
$\log\, \rho_c$ &
    1.892 &
    1.918 &
    1.893 &
    1.823 &
    1.841 &
    1.935 &
    1.869 &
    1.911 &
    1.827 &
    1.893 &
    1.892 \\
$\rho_c/\overline{\rho}$ &
       25 &
       40 &
       35 &
       17 &
       20 &
       73 &
       29 &
      114 &
       96 &
      446 &
       83 \\
$r_{cc}/R_e$ & 
   \ldots &
   \ldots &
   \ldots &
    0.138 &
    0.138 &
    0.046 &
    0.090 &
    0.094 &
    0.122 &
    0.054 &
    0.110 \\
$m_{cc}/M$ & 
   \ldots &
   \ldots &
   \ldots &
    0.023 &
    0.031 &
    0.007 &
    0.014 &
    0.082 &
    0.141 &
    0.085 &
    0.074 \\
$r_{ce}/R_e$ & 
    0.678 &
    0.718 &
    0.618 &
    0.718 &
    0.722 &
    0.822 &
    0.746 &
    0.990 &
    0.990 &
    0.718 &
   \ldots \\
$m_{ce}/M$ & 
    0.921 &
    0.968 &
    0.905 &
    0.935 &
    0.950 &
    0.997 &
    0.972 &
    1.000 &
    1.000 &
    0.995 &
   \ldots \\
$\log\, T_{ce}$ &
    6.458 &
    6.392 &
    6.467 &
    6.452 &
    6.440 &
    6.104 &
    6.388 &
    4.780 &
    4.771 &
    5.988 &
   \ldots \\
Fig. & 
   \ldots & 
   \ldots & 
     3$a$ & 
     3$b$ & 
     3$c$ & 
   \ldots & 
     3$d$ & 
   \ldots & 
   \ldots & 
     3$e$ & 
     3$f$ \\
\enddata
\tablecomments{Quantities listed (units in parentheses):
total mass, $M$ ($M_\odot$);
rotational parameters, $\alpha$, $\eta$;
dimensionless rotational kinetic energy, $t$;
total angular momentum, $J$ (10$^{50}$ g cm$^2$ s$^{-1}$);
equatorial velocity, $V_e$ ($\mbox{km}\ \mbox{s}^{-1}$);
luminosity, $L$ ($L_\odot$);
mean effective temperature, $T_{\mbox{\emph{eff}}}$ (K);
mean surface gravity, $g_s$ ($\mbox{cm}\ \mbox{s}^{-2}$);
equatorial radius, $R_e$ ($R_\odot$);
polar radius, $R_p$;
maximum (normal) distance from the equatorial plane
to the surface of star, $Z_{\mbox{\emph{max}}}$;
central pressure, $P_c$ ($\mbox{dyn}\ \mbox{cm}^{-2}$);
central temperature, $T_c$ (K);
central density, $\rho_c$ ($\mbox{g}\ \mbox{cm}^{-3}$);
mean density, $\overline{\rho}$;
distance in the equatorial plane from the center
to the top of the convective core, $r_{cc}$,
and to the bottom of the convective envelope, $r_{ce}$;
mass enclosed by the upper bounding surface
of the convective core, $m_{cc}$.,
and by the lower bounding surface
of the convective envelope, $m_{ce}$;
and temperature at the bottom
of the convective envelope, $T_{ce}$ (K).}
\end{deluxetable*}


\subsection{1 $\mathbf{\emph{M}_\sun}$ Models}

In Figure 2, we show how some of the characteristics of the $1\ M_\sun$ 
models depend on the dimensionless rotation parameter $\eta$ for $0 \leq 
\alpha \leq 5$.  The relation between $\eta$ and the physical quantity 
$V_e$, the equatorial rotation speed at the stellar surface, is given in 
panel (a), from which it can be seen that for each of the $\alpha$ 
sequences depicted, $V_e$ increases monotonically with $\eta$.  Along the 
curves with $\alpha < \alpha_t$ in Figure 2, the plotted models span the 
range from $\eta = 0$ (i.e., nonrotating) to the value $\eta = 1 + 
\alpha^2$ for which $\Omega_e = \Omega_{cr}$.  Along the curves with 
$\alpha > \alpha_t$, the last plotted model is located adjacent to the 
boundary of the forbidden region in the $(\alpha , \eta)$ plane; for these 
differentially rotating models, the centrifugal and gravitational forces 
have nearly equal magnitudes in the core of the star.  We note that the 
degree of differential rotation increases with $\alpha$, in the sense that 
the configurations corresponding to larger values of $\alpha$ have a 
greater difference between the axial and surface equatorial rates of 
rotation (see \S 2.1).

As panel (b) of Figure 2 makes evident, the radiative luminosities of 
these models are diminished relative to the luminosity $L_0$ of a 
nonrotating $1\ M_\sun$ star.  This is a well-known consequence of 
including rotation in the determination of the equilibrium stellar 
structure (see, e.g., Clement 1979; Bodenheimer 1971).  In the results 
shown in Figure 2, the reduction in $L$ is larger for differentially 
rotating models than it is for models that are uniformly or nearly 
uniformly rotating.  A model with $\alpha = 0$ rotating at the break-up 
rate ($\eta = 1$) has $L/L_0 = 0.78$, while an $\alpha = 2$ model with 
$\eta = 2.42$ has $L/L_0 = 0.15$, a reduction of more than a factor of 6 
from the nonrotating value.  Much of the reason for this behavior lies in 
the effect of rotation on the thermodynamic conditions in the deep, 
energy-producing regions of the stellar interior. For these $1\ M_\sun$ 
models, the contribution of the centrifugal force to supporting material 
against gravity enables the star to emulate an object of lower mass with 
correspondingly reduced values of $P_c$, $T_c$, and $\rho_c$ (e.g., 
Sackmann 1970).  The results presented in panel (c) illustrate the 
dependence of $T_c$ on model rotational properties; similar variations are 
found for both $P_c$ and $\rho_c$.  For rigidly rotating configurations, 
this centrifugal support is largest in the outermost layers of the 
interior, which contain only a small fraction of the stellar mass; in this 
case, $P_c$, $T_c$, and $\rho_c$ are little changed from the values 
appropriate to a nonrotating star of the same mass.  For the $\alpha = 0$, 
$\eta = 1$ model noted previously, $P_c / P_{c0} = 0.94$, $T_c / T_{c0} = 
0.96$, and $\rho_c / \rho_{c0} = 0.98$, where the subscript 0 indicates 
the nonrotating value.  Alternatively, in models for higher values of 
$\alpha$, the effects of rotation are increasingly concentrated toward the 
central regions of the star, with the result that the perturbations to the 
central thermodynamic quantities can be more substantial; for $\alpha = 
2$, $\eta = 2.42$, $P_c / P_{c0} = 0.56$, $T_c / T_{c0} = 0.68$, and 
$\rho_c / \rho_{c0} = 0.81$.  Panels (b) and (c) also indicate that the 
magnitudes of the changes in $L$, $T_c$, and other quantities depend on 
the assumed profile of internal differential rotation.  The model for 
$\alpha = 5$, $\eta = 4$ has $L/L_0 = 0.54$, with $P_c / P_{c0} = 0.72$, 
$T_c / T_{c0} =0.84$, and $\rho_c /\rho_{c0} =0.85$, smaller reductions 
relative to the nonrotating model than those for $\alpha = 2$, $\eta = 
2.42$.  This behavior is an outgrowth of the structural modifications 
arising from the centrifugal force distributions associated with the 
different rotation profiles.  In the $\alpha = 5$ model, the ratio of the 
centrifugal to gravitational force in the equatorial plane, $\Omega^2 r / 
g$ ($r$ is the radial coordinate in the equatorial plane), is sharply 
peaked in the innermost portion of the stellar core, with maximum value 
0.83 at the center, decreasing to $\approx 0.1$ at $r / R_e = 0.3$.  For 
the shallower angular velocity profile of the $\alpha = 2$ model, the 
force ratio decreases from a smaller central value of 0.42 to 0.23 at the 
stellar surface, 10 times the value found throughout the outer 50\% of the 
interior of the $\alpha = 5$ model.

In this connection, we note that Bodenheimer (1971) found that the 
luminosities of models for rotating 30 $M_\sun$ stars, computed assuming a 
variety of internal rotation laws, depended primarily on the total angular 
momentum content of a given stellar model and not on the details of its 
distribution within the interior. Specifically, his results indicated that 
the luminosities of models corresponding to four different prescribed 
distributions of the angular momentum per unit mass decreased with 
increasing total angular momentum $J$, with the relation between $L$ and 
$J$ nearly the same for each model sequence.  If the luminosities of our 1 
$M_\sun$ models are plotted versus their respective $J$ values, an 
analogous reduction in $L$ for increasing $J$ can be discerned.  However, 
the relation between $L$ and $J$ is roughly independent of $\alpha$ only 
for small $J$, and exhibits a clear dependence on $\alpha$ that becomes 
increasingly pronounced as $J$ is made larger.  That the luminosity can be 
even approximately expressed as a function of $J$ necessarily reflects the 
modifications produced by rotation to the thermodynamic conditions in the 
energy-producing core of the star, as described previously (see also Mark 
1968).  The nature and origin of the relation between $L$ and $J$ for 
stars with masses $1 - 2\ M_\sun$ and higher, including its dependence on 
the structural characteristics of the models, will be addressed in detail 
in a subsequent paper in this series.

As in Paper I, we define an average effective temperature through the 
relation $T_{eff} = \left( L / \sigma A \right)^{1/4}$, where $\sigma$ is 
the Stefan-Boltzmann constant and $A$ the area of the stellar surface. The 
results for $T_{eff}$ depicted in panel (d) exhibit dependences on 
$\alpha$ and $\eta$ that are similar to those seen for the quantities 
plotted in the preceding panels.  Reductions by more than $1500\ K$ from 
the nonrotating value ($T_{eff} = 5540\ K$) are possible as, for example, 
in the case $\alpha = 1.5$, $\eta = 2.18$, for which $T_{eff} = 3880\ K$. 
The rotation-induced variations in $T_{eff}$ represent the combined 
effects of changes in both $L$ and $A$.  An indication as to the behavior 
of $A$ can be gleaned from examination of the influence of rotation on the 
stellar size and shape.  This information is presented in panels (e) and 
(f), where we show, respectively, the equatorial radius $R_e$ (in units of 
the radius $R_0$ of the nonrotating, spherical $1\ M_\sun$ model), and the 
polar radius $R_p$ together with $Z_{max}$, the maximum perpendicular 
distance from the equatorial plane to the stellar surface.  These latter 
two quantities are both given as fractions of $R_e$; values of $R_p / R_e$ 
that are $< 1$ reflect a rotational flattening of the configuration, while 
values of $Z_{max} / R_e$ that are $\neq R_p /R_e$ are indicative of a 
deviation from a convex spheroidal shape through the development of a 
concavity in each of the two polar regions of the star.

\begin{figure*}
\includegraphics[angle=270,width=\textwidth]{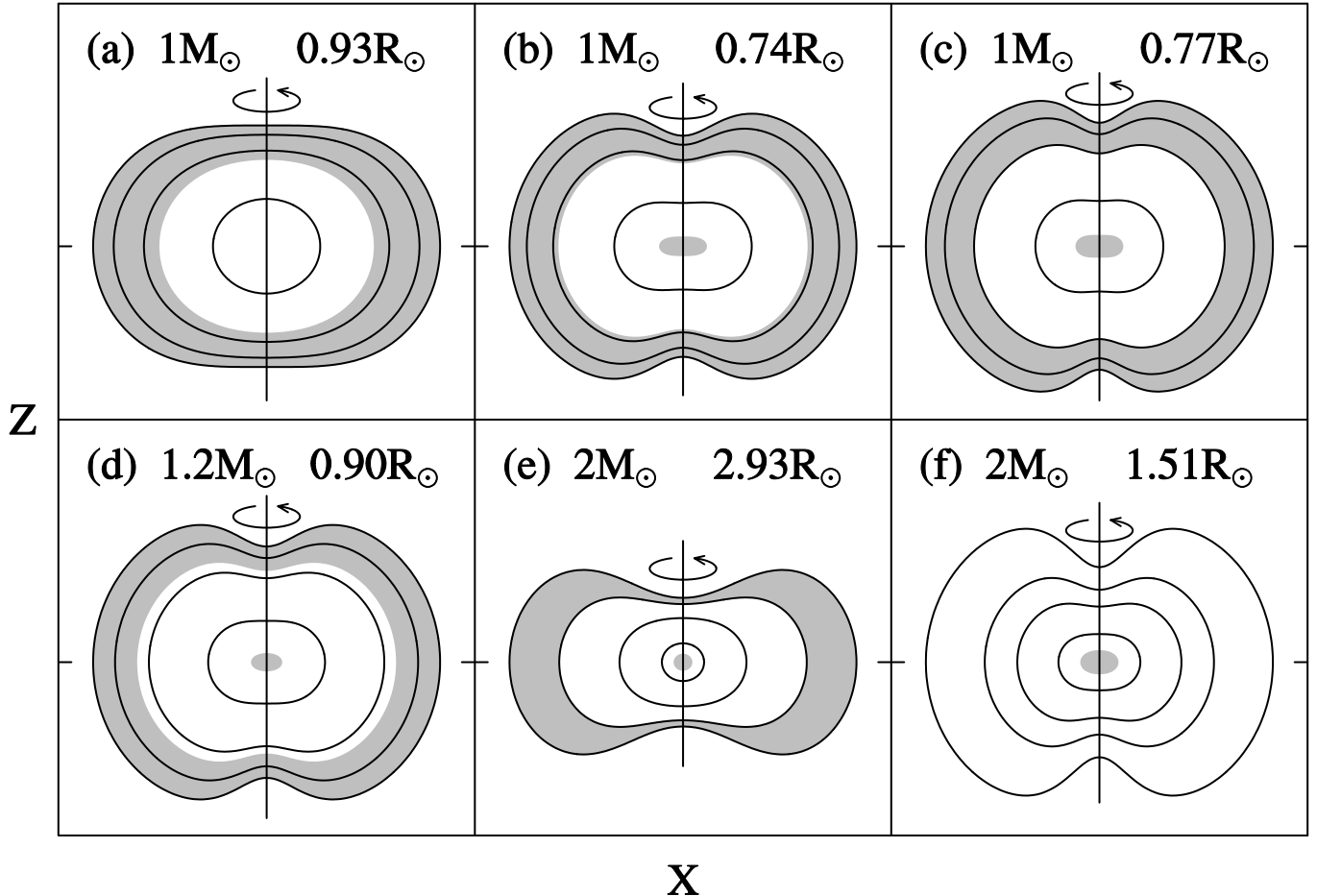}
\caption{Contours of level surfaces in the meridional plane for some of 
the nonspherical models listed in Table 1.  The six rotating models shown 
are defined by the total mass and the two rotational parameters ($M$, 
$\alpha$, $\eta$) as follows: (a) $1\ M_\sun$, 1.5, 1.55; (b) $1\ M_\sun$, 
3.75, 3.58; (c) $1\ M_\sun$, 5, 4.15; (d) $1.2\ M_\sun$, 4, 3.74; (e) $2\ 
M_\sun$, 3, 5.64; and, (f) $2\ M_\sun$, 4.75, 5.9.  From the surface 
inward, the level surfaces depicted in each panel enclose a fraction of 
the total mass equal to 1.000, 0.995, 0.950, and 0.500, respectively.  
The fractional radii in the equatorial plane of these level surfaces for 
the various models are: (a) 1.00, 0.88, 0.71, 0.31; (b) 1.00, 0.90, 0.75, 
0.40; (c) 1.00, 0.89, 0.72, 0.37; (d) 1.00, 0.87, 0.68, 0.34; (e) 1.00, 
0.71, 0.37, 0.12; and, (f) 1.00, 0.66, 0.47, 0.24.  Radiative portions of 
the interior are indicated in white, and convective regions are shaded 
gray.  The fractional equatorial radii and enclosed masses for the 
interfaces between radiative and convective zones in the models are listed 
in Table 1.  The numbers at the tops of the panels denote the total mass 
$M$ and equatorial radius $R_e$ of each model.}
\end{figure*}

Note that the response of the equatorial radius to increasing $\eta$ 
differs greatly depending upon whether $\alpha$ is $\la 1.5$ or $\ga 2$.  
In the former case, the centrifugal force attains its largest value 
relative to the gravitational force at the stellar surface, consequently 
producing a distension of the outer, equatorial layers of the stellar 
interior and an overall increase in $R_e$.  Although $R_p$ decreases 
somewhat relative to the radius of the corresponding nonrotating model, 
the net effect of the changes in $R_e$ and $R_p$ is usually an increase in 
the volume of the rotating star.  For some models with $\alpha \approx 
\alpha_t$, the decrease in $R_p$ can more than compensate for the the 
increase in $R_e$, and the volume of the rotating star is reduced. In the 
case where $\alpha \ga 2$, both $R_e$ and the stellar volume shrink with 
increasing $\eta$ along a constant-$\alpha$ sequence.  For these models, 
the centrifugal force is largest in comparison to the gravitational force 
in the central region of the stellar core, causing the central 
thermodynamic conditions, luminosity, and size (i.e, radius, surface area, 
volume) to assume values that are characteristic of nonrotating stars of 
lower mass.  This raises the possibility that a differentially rotating 
star can imitate a less massive nonrotating star in radius and effective 
temperature, as well as in luminosity.  In panel (f), it can be seen that 
for $\alpha \la 1.5$, the curves for $Z_{max} / R_e$ (solid lines) and 
$R_p / R_e$ (dotted lines) are coincident and $<1$, implying that the 
photospheric shape of these models is oblate spheroidal. Such a model 
($\alpha = 1.5$, $\eta = 1.55$) is depicted in panel (a) of Figure 3.  
For $\alpha > 2$, however, $Z_{max} > R_p$, symptomatic of the development 
of a ``dimple'' or indentation at either pole, as in the cases of the 
models shown in panels (b) ($\alpha = 3.75$, $\eta = 3.58$) and (c) 
($\alpha = 5$, $\eta = 4.15$) of Figure 3.

Panels (g) and (h) of Figure 2 contain results pertaining to the location, 
extent, and properties of convective regions in the models.  In the 
absence of rotation, the internal structure of a $1\ M_\sun$ star consists 
of an inner, radiative core that encompasses $\approx 72$\% of the stellar 
radius and contains $\approx 97$\% of the stellar mass, surrounded by an 
outer, convective envelope.  As is apparent in panel (g), the inclusion of 
rotational effects can modify this basic morphological picture in two 
ways: either by increasing the size of the convective envelope or by 
promoting the formation of a convective core.

For lower values of $\alpha$ (i.e., $\alpha < 3$ in panel [g]), the 
rotationally induced changes in internal structure cause the outer 
convection zone to deepen as $\eta$ increases.  For example, in the model 
with $\alpha = 1.5$, $\eta = 1.55$ (see panel [a] in Figure 3), the radius 
in the equatorial plane of the base of the convective envelope is $r_{ce} 
/ R_e = 0.618$, significantly deeper than the the base radius $r_{ce} / 
R_e = 0.718$ in the nonrotating model.  Associated with this reduction in 
$r_{ce}$ is a decrease in the mass of the core ($m_c / M \approx 0.90$), 
and an increase in the temperature $T_{ce}$ at the bottom of the envelope, 
as can been seen in panel (h); for the $\alpha = 1.5$, $\eta = 1.55$ 
model, $T_{ce} = 2.93 \times 10^6\ K$, as compared with $T_{ce} = 2.46 
\times 10^6\ K$ for the nonrotating model.  Since some chemical species 
(e.g., Li, Be, and B) can be destroyed by thermonuclear reactions at 
temperatures $\ga 2.5 \times 10^6\ K$, enhancements of $T_{ce}$ of this 
magnitude are likely to have consequences for the surface abundances of 
these elements.  In the case of the strongly differentially rotating 
models with $\alpha \geq 3$, the largest centrifugal effects are 
concentrated in the innermost portion of the core, so that the fractional 
thickness of the outer envelope is little affected.  However, as the 
magnitude of the centrifugal-to-gravitational force ratio in the core 
increases for larger $\eta$, the decreasing pressure gradient implied by 
the requirement of hydrostatic equilibrium forces the otherwise stably 
stratified central regions of the interior to become convective.  Under 
these conditions, with the presence of a convective core, the structure of 
the deep interior resembles that of a higher-mass star.  The models shown 
in panels (b) and (c) of Figure 3 are examples of $1\ M_\sun$ stars with 
convective cores; in each of these models, the radial extent of this 
region in the equatorial plane is about 14\% of $R_e$ and contains 
$\approx 2-3$\% of the stellar mass.  For comparison, the convective core 
in a model for a nonrotating $2\ M_\sun$ star has a radius that is about 
12\% of $R_e$ and contains $\approx 14$\% of the stellar mass.

\begin{figure}
\plotone{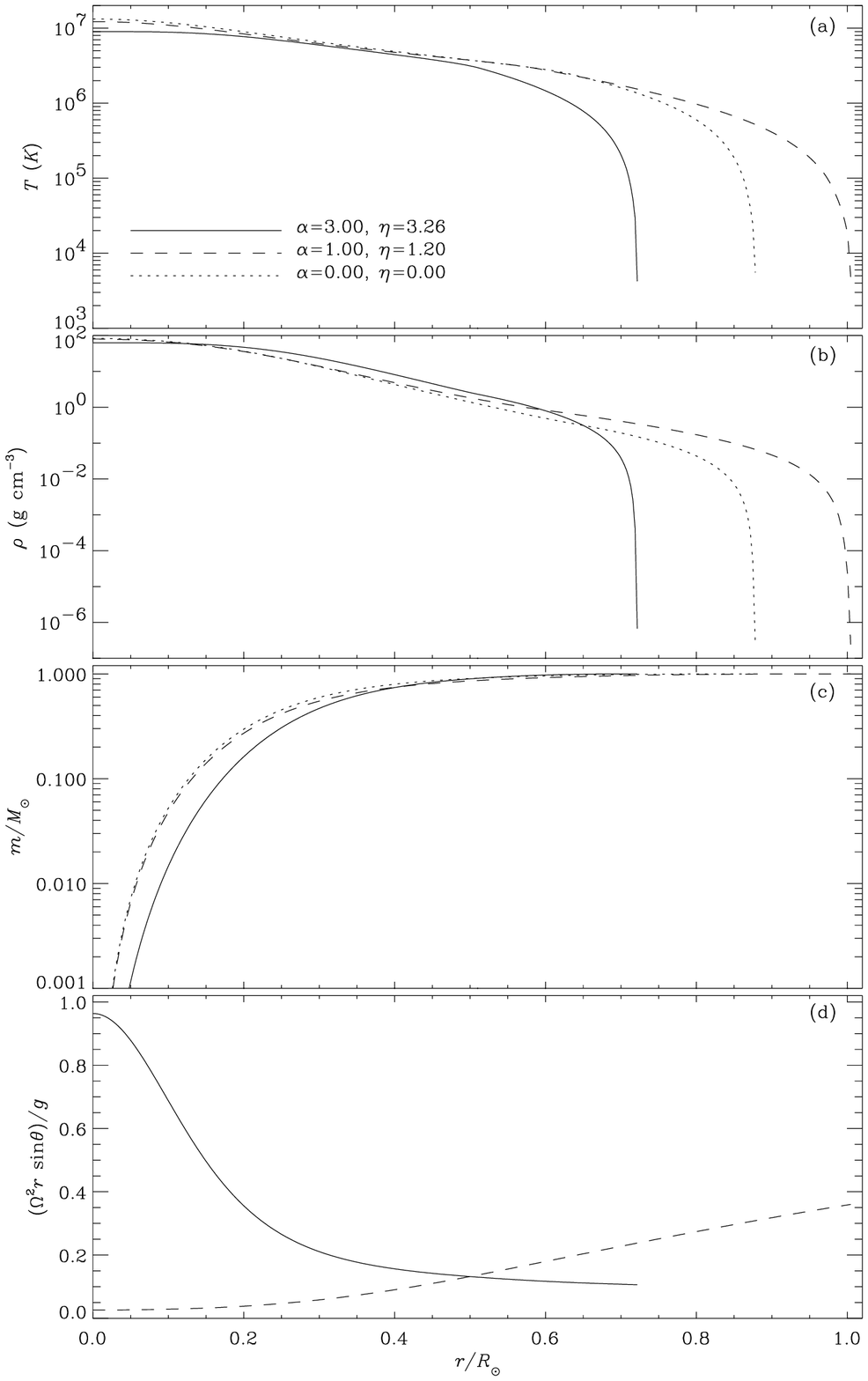}
\caption{Profiles of selected physical quantities in the interiors of $1\ 
M_\sun$ models with $( \alpha, \eta ) = (3.00, 3.26)$ (solid curves), 
$(1.00, 1.20)$ (dashed curves), and $(0.00, 0.00)$ (dotted curves).  The 
profiles depict the dependence of each quantity on radial position $r$ in 
the equatorial plane, from the center of the star to the surface.  The 
various panels show (a) the temperature $T$, (b) the mass density $\rho$, 
(c) the fraction $m/M$ of the total mass contained within a level surface 
that intersects the equatorial plane at $r$, and (d) the ratio $( \Omega^2 
r /g)$ of the centrifugal force to the gravitational force.}
\end{figure}

To illustrate the structural differences between differentially and 
near-uniformly rotating stars, in Figure 4 we show the profiles of several 
physical quantities for models with $( \alpha, \eta ) = (3.00, 3.26)$, and 
$(1.00, 1.20)$, along with the corresponding results for the nonrotating 
model.  The luminosities and total angular momenta of the two rotating 
models are $L / L_0 = 0.201$ and $J = 9.23$ ($\alpha = 3$), and $L / L_0 = 
0.634$ and $J = 5.30$ ($\alpha=1$), respectively, where $J$ is meaasured 
in units of $10^{50}$ g cm$^2$ s$^{-1}$. In panels (a) and (b), the 
temperature $T$ and mass density $\rho$ are depicted as functions of the 
radial position $r$ (measured in units of the present-day solar radius 
$R_\sun$) in the equatorial plane of the star. The central values of both 
quantities exhibit rotation-induced reductions relative to the nonrotating 
case, the magnitudes of these modifications being larger for $\alpha = 3$ 
($T_c / T_{c0} = 0.68,\ \rho_c / \rho_{c0} = 0.76$) than for $\alpha = 1$ 
($T_c /T_{c0} = 0.92,\ \rho_c /\rho_{c0} = 0.97$).  As noted previously, 
this behavior is a consequence of differences in the magnitudes and 
distributions of the centrifugal force in the two rotating models.  These 
distinctions can be clearly seen in the profiles of the 
centrifugal-to-gravitational force ratio shown in panel (d).  For $\alpha 
= 3$, the force ratio is largest in the deep interior, attaining a maximum 
value of 0.96 at the stellar center and decreasing outwards to a magnitude 
$\approx 0.1$ in the photosphere. For $\alpha = 1$, the ratio increases 
monotonically throughout the interior, rising from a central value of just 
0.026 to a maximum of 0.36 at $R_e$. The substantial contribution of the 
centrifugal force to the support of the innermost regions of the $\alpha 
=3$ model is responsible for the considerable enhancement of the density 
scale height there, evident in panel (b); $\rho$ declines by only $\approx 
10$\% over the inner 20\% of the stellar interior.  The resulting changes 
in the internal mass distribution (panel [c]) lead to a star with a 
smaller radius, $R_e = 0.72\ R_\sun$ as opposed to $R_e = 0.88\ R_\sun$ 
for the nonrotating model.  Alternatively, for the $\alpha = 1$ model, the 
lack of centrifugal support in the core of the star leads to temperature, 
density and mass distributions therein that closely resemble those of the 
nonrotating model. Closer to the surface, however, the density 
distribution becomes extended, a product of the increasing centrifugal 
reduction of gravity in the outer layers of the interior; as a result, the 
stellar radius is larger than that of the nonrotating model, $R_e = 1.00\ 
R_\sun$.

\begin{figure}
\plotone{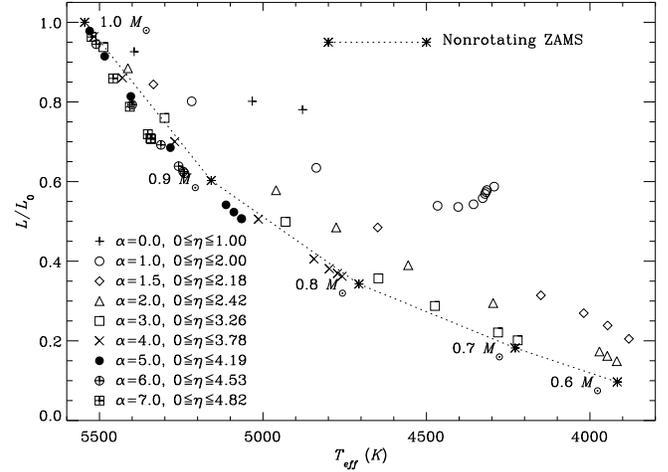}
\caption{A theoretical HR diagram showing the positions of models for 
rotating, ZAMS, $1\ M_{\sun}$ stars.  Luminosities are given in units of 
the luminosity $L_0 = 0.653\ L_\sun$ of a non-rotating, $1\ M_\sun$ star.  
The various symbols denote models constructed using the rotation law of 
equation (1) with the values of the parameters $(\alpha, \eta)$ listed in 
the Figure.  The ZAMS for non-rotating stars is indicated by the dotted 
line, with the positions of models for masses 0.6, 0.7, 0.8, 0.9, and 1.0 
along it marked by an $\ast$ symbol.}
\end{figure}

\begin{figure}
\plotone{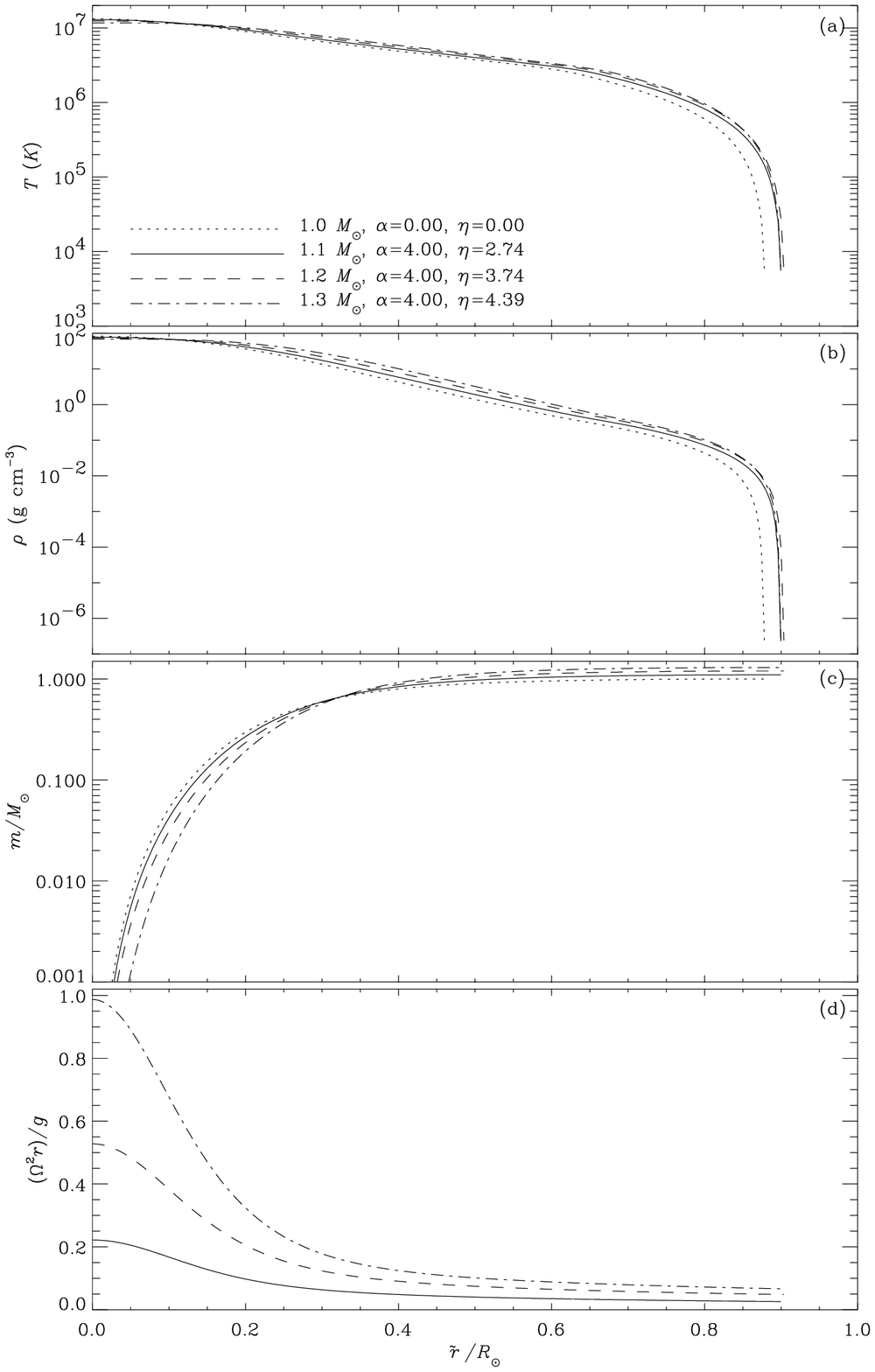}
\caption{Profiles in the equatorial plane of the temperature (a), density 
(b), mass fraction (c), and the ratio of the centrifugal to gravitational 
acceleration (d) for a nonrotating, $1\ M_\sun$ model ($\alpha = \eta = 
0$) and three \emph{solar look-alike} models. The latter models were 
obtained for $(M, \alpha, \eta) = (1.1\ M_\sun, 4.00, 2.74),\ (1.2\ 
M_\sun, 4.00, 3.74)$ and $(1.3\ M_\sun, 4.00, 4.39)$, and have values of 
$T_{eff}$ and $L$ that differ from those of the non-rotating, solar model 
by about 1\% or less.}
\end{figure}

\subsection{Solar Look-Alike Models}

Figure 5 is a theoretical HRD for $1\ M_\sun$ models that span a broad 
range of internal rotational states, from uniform rotation to extreme 
differential rotation (up to $\Omega_0 = 50\ \Omega_e$ for $\alpha = 7$). 
The nonrotating ZAMS is delineated by a dotted line, with the positions of 
several specific models for $0.6 \leq M \leq 1.0\ M_\sun$ indicated. 
Clearly, as already implied by panels (b) and (c) of Figure 2, the 
locations of rotating models in such a diagram are displaced to the right 
of and below the position they would occupy in the absence of rotation, 
toward lower values of both the luminosity and the effective temperature.  
Models with uniform or near-uniform internal rotation (i.e., $\alpha = 0,\ 
1$) lie well above the nonrotating ZAMS, while those for which the degree 
of differential rotation is substantial ($\alpha \ge 3$) have positions in 
the HRD that collectively approximate the nonrotating ZAMS over virtually 
the entire mass range depicted.  For example, the rotating $1\ M_\sun$ 
model ($\alpha = 3.75$, $\eta = 3.58$) shown in panel (b) of Figure 3 is 
characterized by $T_{eff} = 4700\ K$ and $L /L_\sun = 0.224$, which is 
indistinguishable from the values $T_{eff} = 4710\ K$, $L / L_\sun =0.224$ 
for a nonrotating, $0.8\ M_\sun$ model.  Such a coincidence of the 
positions of stars of different mass in the classical HRD is made possible 
by the fact that the largest structural changes take place in the cores of 
strongly differentially rotating models.  Specifically, the decrease in 
$L$ brought about by the smaller values of $P_c$, $T_c$, and $\rho_c$, 
together with the corresponding reduction in $R_e$, enable such a model to 
effectively mimic a lower mass stellar model in which rotational effects 
are not included.

Figure 6 illustrates some of the internal properties of three so-called 
\emph{solar look-alike} models, that is, models for differentially 
rotating stars with masses $> 1\ M_\sun$ that have many physical 
attributes in common with the model for a nonrotating, $1\ M_\sun$ star.  
Inspection of panels (a) and (b) reveals that the profiles and the central 
values of the temperature and density for the 1.1, 1.2, and 1.3 $M_\sun$ 
models depicted therein are quite close to those of the solar-mass model 
with $\alpha = \eta =0$.  A consequence of these structural similarities 
is that many of the general characteristics of the models are nearly 
identical.  For example, the $1.2\ M_\sun$ model which, in the absence of 
rotation, would have $L / L_\sun= 1.591$, $R_e / R_\sun = 1.129$, and 
$T_{eff} = 6110\ K$ instead has $L /L_\sun = 0.653$, $R_e / R_\sun = 
0.904$, and $T_{eff} = 5546\ K$, compared with $L / L_\sun =0.653$, $R_e / 
R_\sun = 0.878$, and $T_{eff} = 5545\ K$ for the nonrotating, $1\ M_\sun$ 
model.  A cross-section in the meridional plane of the rotating $1.2\ 
M_\sun$ interior is shown in panel (d) of Figure 3, from which it can be 
seen that a solar-like convection zone having $r_{ce} / R_e = 0.746$ is 
present in the layers beneath the photosphere.  With a surface equatorial 
rotation speed $V_e = 112$ km s$^{-1}$, such an object could be mistakenly 
identified as a rapidly rotating $1\ M_\sun$ star, if observations were 
analyzed through comparison with nonrotating stellar models.  \emph{That 
the values of measureable or inferrable stellar properties can be the same 
in rotating and nonrotating models for different masses represents a 
potential source of ambiguity in the interpretation of a variety of 
observations.} Panels (c) and (d) reveal subtle differences in the 
internal mass distributions and in the variation of the ratios of the 
centrifugal to gravitational force in the equatorial planes of these solar 
look-alike models.  Future space-based asteroseismological observations 
may be capable of exploiting such differences in internal structure to 
distinguish slowly rotating stars from more rapidly, differentially 
rotating higher-mass stars that happen to have the same values of $L$ and 
$T_{eff}$ and thus occupy the same position in the HRD (see, e.g., Lochard 
et al. 2005).

\subsection{2 $\mathbf{\emph{M}_\sun}$ Models}

In Figure 7, we present a summary of the rotational dependence of $2\ 
M_\sun$ model characteristics, using the same format as adopted for Figure 
2.  The model for a non-rotating star of this mass has $L_0 = 15.47\ 
L_\sun$, $R_0 = 1.59\ R_\sun$, and $T_{eff} = 9090\ K$, and a convective 
core with $r_{cc} / R_e = 0.122$, $m_{cc} / M = 0.141$; such a model has a 
thin, subsurface convective layer, $r_{ce} / R_e = 0.99$, that contains a 
negligible fraction of the stellar mass.  The results shown in Figure 7, 
which span the rotational parameter ranges $0 \leq \alpha \leq 5$, $0 \leq 
\eta \leq 7$, exhibit many of the same behaviors as noted previously in 
connection with the $1\ M_\sun$ models of Figure 2.  In particular, along 
each constant-$\alpha$ sequence, the luminosity (panel [b]), central 
temperature, (panel [c]) and effective temperature (panel [d]) all 
initially decrease as $\eta$ is increased, with the largest reductions in 
these quantities occurring in the cases of differentially rotating models 
with $\alpha \ge 3$.  For $\alpha =$ 1, 2, both $L / L_0$ and $T_c / 
T_{c0}$ pass through minima and then increase slightly with $\eta$ beyond 
those points, whereas $T_{eff}$ continues to decrease with $\eta$ for all 
$\alpha$ sequences.  The central pressure $P_c$ (panel [e]), on the other 
hand, is an increasing function of $\eta$ for $0 \leq \alpha \leq 3$, and 
displays a non-montonic dependence on $\eta$ for the model sequences 
corresponding to $\alpha = 4,\ 5$.  The origin of this variation is the 
mass dependence of $P_c$ for stars on the non-rotating ZAMS.  In our SCF 
models without rotation, $P_c$ has a maximum value for $M$ near $1.6\ 
M_\sun$, and is decreasing for both larger and smaller $M$ values.  The 
maximum occurs for a mass close to the value marking the internal 
structural transition between stars with radiative cores and convective 
envelopes and stars with convective cores and radiative envelopes.  
Hence, insofar as the properties of a rotating star are like those of a 
non-rotating star of lower mass, we expect the $P_c$ values of these $2\ 
M_\sun$ models to increase as the effects of rotation become more 
pronounced.  For cases in which the central thermodynamic conditions are 
significantly perturbed by rapid, differential rotation in the core of the 
star, this mass-lowering effect can be large enough to shift $P_c$ to 
values characteristic of nonrotating stars with $M < 1.6\ M_\sun$ (i.e., 
on the other side of the central pressure peak), as in the $2\ M_\sun$ 
models for $\alpha =$ 4, 5.

\begin{figure}
\plotone{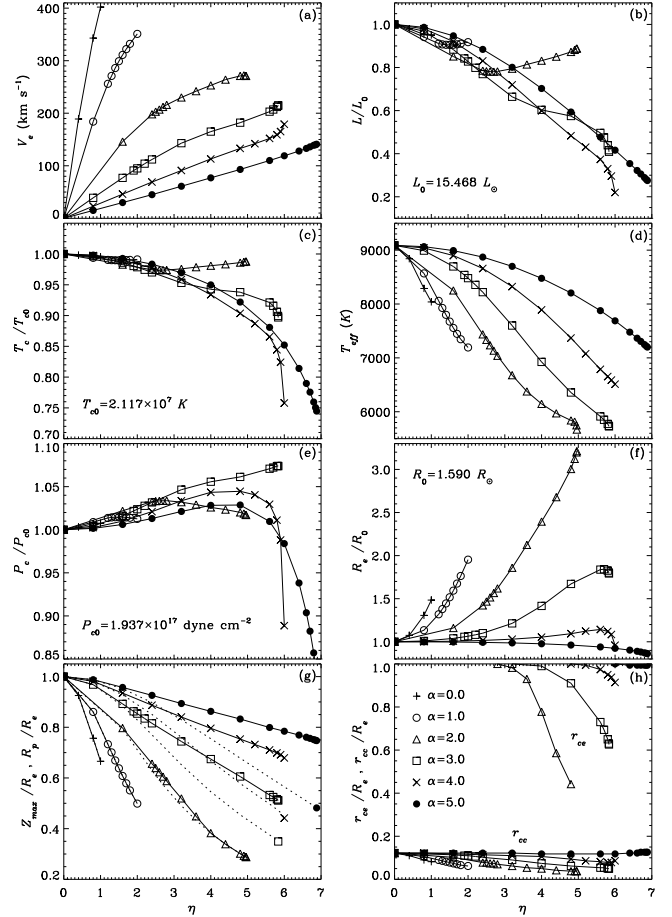}
\caption{Selected properties of differentially rotating, $2\ M_\sun$, ZAMS 
stellar models, as in Figure 2.  The quantities depicted in the various 
panels include: (a) $V_e$, (b) $L$, (c) $T_c$, (d) $T_{eff}$, (e) $P_c$, 
(f) $R_e$, (g) $Z_{max}$ and $R_p$, and (h) $r_{ce}$ and $r_{cc}$.}
\end{figure}

Panels (f) and (g) of Figure 7 convey information pertaining to the 
photospheric sizes and shapes of the models.  Those that rotate nearly 
rigidly are equatorially distended, and, in the direction perpendicular to 
the equatorial plane, have their largest dimension along the rotation axis 
($Z_{max} = R_p$); since $R_p < R_e$, these models (like their $1\ M_\sun$ 
counterparts) have a flattened, spheroidal shape.  Models with increasing 
degrees of differential rotation (i.e., with $\alpha \ga 3$) develop polar 
concavities (as indicated by $Z_{max} > R_p$), and ultimately (i.e., for 
$\alpha = 4,\ 5$ and sufficiently large $\eta$) become more compact, with 
$R_e < R_0$.  Panels (e) and (f) of Figure 3 give cross-sectional 
representations of the photospheric shapes of $2\ M_\sun$ models for 
$(\alpha,\ \eta) = (3.00,\ 5.64)$ and $4.75,\ 5.90)$, repectively; in the 
former model, $R_e = 2.93\ R_\sun > R_0$, while for the latter, $R_e = 
1.51\ R_\sun < R_0$.

The influence of rotation on the occurrence and extent of convective 
regions within the models is explored in panel (h) of Figure 7.  The 
convective core, a salient feature of the non-rotating stellar interior, 
generally decreases in size as the value of $\eta$ is increased.  This 
contraction stems from a reduction in the magnitude of the radiative 
gradient $\nabla_{rad}$ in the core region of the star, a result of the 
way in which the central thermodynamic conditions are modified by rotation 
(see, e.g., MacGregor \& Gilliland 1986); with $\nabla_{rad}$ smaller, the 
size of the region wherein it exceeds the adiabatic gradient $\nabla_{ad}$ 
shrinks accordingly.  Only for the most rapidly, differentially rotating 
models does $r_{cc}$ show a modest increase compared to the non-rotating 
model; in panel [h], the model for $(\alpha,\ \eta) = (5.00, 6.88)$ has 
$r_{cc} / R_e = 0.124$.  In this case, as was seen for the $1\ M_\sun$ 
models having convective cores in Figure 2, the growth of $r_{cc}$ can be 
traced to a \emph{larger} value of $\nabla_{rad}$, produced by a 
centrifugal-to-gravitational force ratio that is nearly unity at the 
stellar center. Note also that for some $2\ M_\sun$ models, the convection 
zone underlying the stellar photosphere can extend into the stellar 
interior by more than the few $10^{-3}\ R_0$ that is the thickness of this 
region in the absence of rotation.  As can be seen, for example, in panel 
(e) of Figure 3, the base of the convective envelope in the equatorial 
plane of the model with $(\alpha,\ \eta) = (3.00, 5.64)$ is located at 
$r_{ce} / R_e = 0.718$, a fractional depth which is the same as in the 
non-rotating $1\ M_\sun$ model.  This leads to the intriguing possibility 
that solar-like oscillations, driven by turbulent convection, may be 
excited in stars that would normally be too massive to generate them. The 
results shown in panel (h) suggest that such a solar-like convective 
envelope is most likely to occur in $2\ M_\sun$ models with intermediate 
differential rotation (say., $\alpha \approx \alpha_t = 2.83$).  For a 
profile of this kind, the inner and outer portions of the interior can 
each rotate rapidly enough to both significantly perturb the thermodynamic 
conditions in the core and increase $R_e$ by extending the stellar 
envelope.

\begin{figure}
\plotone{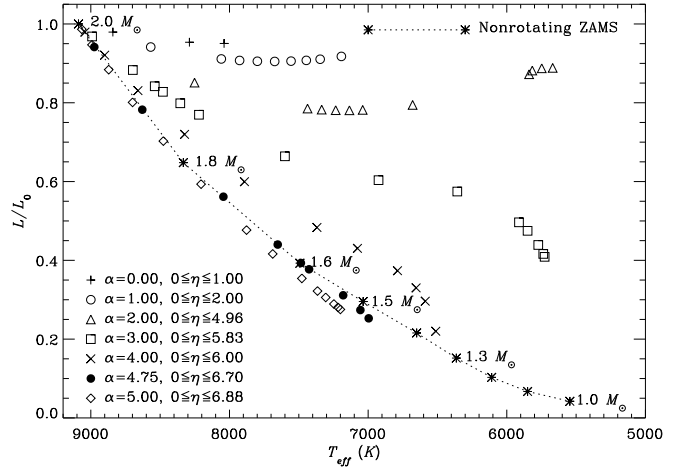}
\caption{A theoretical HR diagram showing the positions of models for 
rotating, $2\ M_\sun$ ZAMS stars, as in Figure 5.  Luminosities are given 
in units of the luminosity $L_0 = 15.468\ L_\sun$ of a nonrotating, $2\ 
M_\sun$ star.  The ZAMS for nonrotating stars is indicated by the dotted 
line, with the positions of models for masses 1.0, 1.1, 1.2, 1.3, 1.4, 
1.5, 1.6, 1.8, and 2.0 $M_\sun$ along it marked by an $\ast$ symbol.}
\end{figure}

In Figure 8, we show the positions of models for differentially rotating, 
$2\ M_\sun$ stars in a theoretical HRD.  As in Figure 5, the location of 
the non-rotating ZAMS is indicated by a dotted line, with the positions of 
models for stars with masses in the range $1.0 \leq M \leq 2.0\ M_\sun$ 
indicated along it.  This HRD for rotating $2\ M_\sun$ models has a number 
of features in common with the corresponding representation of $1\ M_\sun$ 
properties.  Without exception, models are shifted from the non-rotating 
location to new positions characterized by lower $T_{eff}$ and $L$ values.  
For $\alpha \la 3$, these new positions lie to the right-hand side of 
(i.e., above) the non-rotating ZAMS; for rotation that is increasingly 
differential, however, the model locations approach the non-rotating ZAMS, 
moving just to the left-hand side of (i.e., below) it for $\alpha = 5$.  
For $\alpha = 4.75$, the model positions are distributed along the 
non-rotating ZAMS, delineating its track in the HRD for masses greater 
than about 1.5 $M_\sun$.  This coincidence again raises the possibility 
that a rapidly, differentially rotating $2\ M_\sun$ star could effectively 
masquerade as a non-rotating star of lower mass.  As an example, with $L / 
L_\sun = 6.083$, $R_e / R_\sun = 1.512$, and $T_{eff} = 7480\ K$ the model 
for $(\alpha,\ \eta) = (4.75, 5.90)$ shown in panel (f) of Figure 3 
closely resembles a non-rotating $1.6\ M_\sun$ star, for which $L / L_\sun 
= 6.075$, $R_e / R_\sun = 1.466$, and $T_{eff} = 7490\ K$.
 
\section{Summary and Discussion}

The SCF approach to computing the structure of rotating stars is robust 
and efficient, capable of yielding converged models of stars of all 
masses, with internal angular-velocity distributions covering the range 
from uniform to extreme differential rotation.  The code described in 
Paper I has been upgraded throught the introduction of opacities, energy 
generation rates, and an equation of state that are close to 
state-of-the-art, and with the introduction of a mixing-length treatment 
of convective energy transport.  We have used this revised code to conduct 
a systematic study of the properties of models for rotating ZAMS stars 
with masses $1 \leq M \leq 2\ M_\sun$, assuming that the rotation of the 
stellar interior can be adequately represented by the parameterized 
angular-velocity profile given in equation (1).

Our results suggest that rotation affects virtually every characteristic 
of the intermediate- and low-mass stars considered herein, with the 
magnitude, and sometimes even the sense, of the changes in fundamental 
attributes depending on both the rate and degree of the differential 
rotation in the stellar interior.  For models that are uniformly or nearly 
uniformly rotating, the centrifugal contribution to the hydrostatic 
support of the star is largest in the outer layers of the interior; for 
models that are strongly differentially rotating, the centrifugal force is 
largest compared to gravity in the core of the star.  Luminosities and 
effective temperatures are always diminished relative to nonrotating stars 
of the same mass, while equatorial radii increase in uniformly rotating or 
moderately differentially rotating models, and decrease in models where 
the difference between the angular velocity at the center and that at the 
equator is sufficiently large.  As the degree of differential rotation is 
increased, the photospheric shape of the star changes from a convex 
surface closely approximating a spheroid flattened along the rotation axis 
to a roughly oblate surface with deepening polar concavities; in such 
cases, the greatest perpendicular distance from the equatorial plane to 
the surface occurs away from the axis of rotation, and is always less than 
the equatorial radius of the star.  In $1\ M_\sun$ models, the effects of 
rotation can either increase the thickness of the outer convective 
envelope, or contribute to the formation of a convective core.  In 
rotating $2\ M_\sun$ models, the size of the convective core is diminished 
relative to that found in the absence of rotation, and an extensive, 
solar-like convection zone can be present in the outer layers of what 
would otherwise be a stable radiative envelope.

The extent to which the results of these computations accurately depict 
the stellar structural modifications arising from rotation depends on the 
validity of a number of simplifying assumptions and approximations made in 
the course of developing the basic model, as well as on the inherent 
limitations of the SCF method itself.  Here, we briefly address the most 
salient of the factors that might restrict the applicability of these 
results, noting where improvements and extensions can (or cannot) be made 
(see also Paper I).

1). The modified SCF iterative scheme utilized in the present 
investigation necessarily requires the angular velocity distribution 
within the star to be conservative, and thus expressible as a function of 
just the perpendicular distance from the rotation axis, $\Omega = 
\Omega(\varpi)$ (see eq. [1]).  Although this specification facilitates 
the construction of models through the considerable mathematical 
simplification it introduces, its effect on some computed stellar 
structural characteristics, such as the photospheric shapes and core 
physical properties of very rapidly, differentially rotating models, is 
likely to differ from that produced by a rotation law of (say) the form 
$\Omega = \Omega(r)$. There is some evidence, from both simulations and 
observations, for the occurrence of differential rotation of the type 
given by equation (1) (see, e.g., Dobler, Stix, \& Brandenburg 2006, and 
references therein).  Yet analyses of solar acoustic oscillations yield a 
picture of the large-scale internal dynamics of the Sun which is not in 
accord with the angular velocity being constant on cylindrical surfaces 
(see Thompson et al. 2003).  We note that use of the SCF method precludes 
consideration of non-conservative rotation laws (e.g., $\Omega = 
\Omega(r)$); however, detailed treatment of the hydrodynamical and 
magnetohydrodynamical processes affecting the internal rotational states 
of the Sun and stars is presently beyond the scope of $any$ of the extant 
structural/evolutionary models.

2). The model treats only the prescribed rotational motion of the stellar 
interior, omitting any meridional circulatory flow and its consequent 
effects on the internal angular momentum distribution.  For models with 
radiative envelopes, rotationally driven circulation will cause the 
rotation profile to deviate over time from the state given by equation 
(1), unless such evolution is mitigated by additional angular momentum 
transport mechansims (see, e.g., Maeder \& Meynet 2000, and references 
therein).  For models with convective envelopes, the rotation profile in 
the outer layers of the interior is the product of the complex interplay 
between meridional circulation and turbulent heat and angular momentum 
transport (Rempel 2005; Miesch, Brun, \& Toomre 2006); whether $\Omega$ is 
solar-like or constant on cylindrical surfaces depends on the latitudinal 
entropy distribution in the subadiabatically stratified layers below the 
convection zone.

3). The model employs a simplified treatment of convection, locating 
unstable regions by application of the Schwarzschild criterion and 
utilizing an averaged, rotationally modified mixing-length description 
(see \S2.2) to determine the structure of an outer convective envelope, if 
present.  Use of the Solberg-H\o iland condition (e.g., Ledoux 1965; 
Kippenhahn \& Weigert 1990) to ascertain the onset of convective 
instability would account for the direct influence of the centrifugal 
force. Since models computed for the rotation law of equation (1) have 
$\partial j / \partial \varpi >0$, where $j = \Omega \varpi^2$ is the 
specific angular momentum (see Paper I), this change would likely decrease 
the equatorial-plane thickness of an outer convection zone, and produce a 
latitudinal variation in $r_{ce}$.  However, for rapid rotation, there is 
considerable uncertainty regardless of the convection criterion/model 
adopted, because of both the rudimentary nature of extant treatments of 
convective energy transport and the universal assumption of axisymmetry 
among rotating stellar models.

We reiterate that the present model describes a chemically homogeneous, 
ZAMS star, and neglects effects associated with the structural and 
compositional evolution of the stellar interior.  We are presently 
developing a mean-field hydrodynamics-based treatment of turbulent 
chemical and angular momentum transport which maintains conservative 
rotation profiles, thus making it possible to investigate the 
main-sequence evolution of these SCF models.  Despite the shortcomings 
enumerated above, we believe that the basic model and method described 
herein compare favorably with alternative approaches to determining the 
structure and evolution of rotating stars.  The most widely used of these 
(e.g., Meynet \& Maeder 1997) relies on the use of approximate, Roche-like 
equipotentials to represent the internal gravity of the star, thereby 
strictly limiting it to describing slowly rotating stars to ensure the 
accuracy of the computed models.  Whatever its drawbacks, the SCF method 
yields two-dimensional, axisymmetric configurations that represent fully 
consistent solutions to both the set of stellar structure equations {\it 
and} Poisson's equation for the gravitational potential.

The results presented herein have a number of implications for young stars 
with masses between 1 and $2\ M_\sun$. Measurements of projected 
equatorial rotation speeds in excess of 100 km s$^{-1}$ for some 
solar-type stars in young open clusters (see, e.g., Stauffer 1991) raise 
the possibility that the structure and properties of these objects could 
be significantly altered from those of nonrotating stars of the same mass. 
Such stars, provided that their interior rotation is sufficiently 
differential, could in actuality be somewhat more massive objects for 
which surface rotation as rapid as that indicated by observations is the 
normally expected ZAMS state. The question of whether or not the kind of 
strong differential rotation required to produce such ambiguity is present 
within the interiors of some low- and intermediate-mass stars may 
ultimately be resolved through space-based asteroseismological 
observations, which should allow low-resolution inversions of the rotation 
profiles in the inner $\sim30\%$ of the stellar radius (see Gough \& 
Kosovichev 1993). Asteroseismology from space may also afford the means 
for identifying individual stars whose rotation enables them to pose as 
lower-mass objects, since, when effects associated with asphericity are 
not too large, the average mode frequency spacing is sensitive to the mean 
density, a quantity which Table 1 reveals to be different for 
\emph{look-alike} models.

If rapid differential rotation is a possibility for objects in the mass 
range spanned by the models of \S3, then the associated changes in 
structure and properties could have consequences for a variety of 
important astrophysical processes that take place within and around such 
stars. The reduced radiative luminosity of a rapidly rotating young Sun 
would likely influence the evolution of the solar nebula, and further 
exacerbate the discrepancies between the properties of nonrotating models 
and observational inferences indicating a higher ZAMS luminosity (see, 
e.g., Sackmann \& Boothroyd 2003). The rotationally induced deepening of a 
sub-photospheric convection zone, together with the increase in 
temperature of the material at the base, could contribute to the depletion 
of lithium in the stellar surface layers by reducing the thickness of the 
region through which chemical species must be transported in order to be 
destroyed by nuclear processes. The formation of a solar-like convective 
envelope in a young, differentially rotating, $2\ M_\sun$ star could 
excite global oscillations, and be accompanied by the operation of a 
solar-like hydromagnetic dynamo. Dynamo-generated fields that diffuse into 
and are retained by the radiative interior (see, e.g., Dikpati, Gilman, \& 
MacGregor 2006) could enable the star to remain magnetic long after 
spin-down and the elimination of nonuniform rotation have led to the 
disappearance of both the surface convective layer and the dynamo. Each of 
these possibilities will be addressed in forthcoming papers.

\acknowledgments
We wish to thank D.~A.\ VandenBerg for providing software from his 
stellar-evolution code that was used to handle the input physics in our 
SCF code, and we wish to thank J.~Christensen-Dalsgaard for the use of his 
stellar-evolution code to help us validate our models. This work was 
supported in part by an Astronomy \& Astrophysics Postdoctoral Fellowship 
under award AST-0401441 (to T.~S.~M.) from the National Science 
Foundation.

\end{document}